**A Novel Nb-Based Eutectic Superalloy with Exceptional Ultrahigh-Temperature Mechanical Properties**

Ayeman M. Nahin[1,†], Alfredo Navarrete[2,†], Xiaokun Yang[1], Eric A. Lass[2,*], Mingwei Zhang[1,*]

[1]Department of Materials Science and Engineering, University of California, One Shields Ave., Davis, CA 95616, USA

[2]Department of Materials Science and Engineering, University of Tennessee, Knoxville, TN 37996, USA

[†] These authors contribute equally to this paper.

*Corresponding Authors:

Eric A. Lass, Associate Professor

Department of Materials Science and Engineering

University of Tennessee, Knoxville

Email: elass@utk.edu

Mingwei Zhang, Assistant Professor

Department of Materials Science and Engineering

University of California, Davis

Email: mwwzhang@ucdavis.edu

## Abstract

Refractory alloys operating above 1300 °C, beyond the limit of Ni-based superalloys, must balance high-temperature strength, thermal stability, low density, low cost, and room-temperature tensile ductility, a combination that existing refractory alloys have yet to achieve. Here we report a castable Nb-based eutectic superalloy, Nb-10Mo-9.5C (at. %, NMC-1), comprising a uniform lamellar structure of Nb-Mo solid solution and $Nb_2C$ carbide phases. NMC-1 achieves 0.2% yield strengths of ~300 MPa and ~200 MPa at 1300 °C and 1500 °C, respectively, among the highest reported for Nb-based alloys, with pronounced strain hardening, non-zero room-temperature tensile ductility, and no observed microstructural coarsening after 100 h at 1400 °C. Its low density (8.65 g/cc) and cost (~$85/kg) yield a specific-strength-per-cost merit index that far surpasses commercial Nb, Mo, Ta, and W alloys, establishing a new design paradigm for ultrahigh-temperature structural materials.

Advances in aerospace propulsion demand structural materials exceeding Ni-based superalloys' maximum service limit of 1300 °C [1,2]. Refractory metals, with their higher melting points, are prime candidates for these extreme environments. Candidate ultrahigh-temperature materials must meet stringent criteria: melting point >2000 °C, high yield strength retention beyond 1300 °C, excellent thermal stability, low density, low cost, and sufficient room-temperature (RT) tensile ductility.

Nb-based alloys are prime candidates for these applications, given niobium's high melting point (2477 °C), excellent ductility, body-centered cubic (BCC) phase stability, low density (8.57 g/cc), and low cost ($88/kg). However, commercial alloys face significant limitations. C-103 (Nb-10Hf-1Ti, wt. %) exhibits modest strength (300 MPa RT, 100 MPa at 1300 ºC [3]) with rising Hf costs. FS-85 (Nb-28Ta-10W-1Zr, wt. %)) and WC3009 (Nb-30Hf-9W, wt. %)) achieve higher high-temperature strength (~150 and 200 MPa at 1300 °C, respectively [4]) but suffer from high density (>10 g/cc) and prohibitive Ta/Hf costs. Nb521 (Nb-5W-2Mo-1Zr-0.01C, wt. %) improves performance through $Nb_2C$ and ZrC dispersion strengthening (350 MPa RT, 250 MPa at 1300 °C) at lower density (8.8 g/cc) and cost (~$600/kg) [5,6], though strength gains remain insufficient. Additive manufacturing can enhance strength up to 1200 °C but loses viability above 1300 °C due to insufficient strength gains and high powder costs (~$2500/kg) [6].

Refractory high-entropy alloys (RHEAs) offer an alternative by replacing single base elements with concentrated solid solutions [7–9]. Single-phase BCC systems (e.g., Mo-Nb-Ta-W-V) display exceptional ultrahigh-temperature compressive strength (400 MPa at 1600 °C [10]) but lack tensile ductility [11], while Hf-Nb-Ta-Ti-Zr systems show extensive RT tensile ductility with insufficient high-temperature strength [12]. Multiphase BCC-B2 refractory high-entropy superalloys (RHSAs) show promise [13–16] but suffer from low B2 solvus temperatures (<1100 °C) [14] and grain boundary embrittlement [17]. Newer Ru-containing [18–20] and Ta-Re-based systems [21] offer higher solvus temperatures with unproven mechanical performance and high cost due to Ru/Re additions. Broadly, multiphase refractory alloys require improved phase stability to 2000 °C, enhanced coarsening resistance above 1300 °C, and minimal grain boundary segregation.

Utilizing eutectic alloys has been demonstrated as a sound materials development strategy for designing alloys with unique elevated temperature mechanical properties. Near-eutectic Al-Ce alloys demonstrate superior elevated-temperature strength retention (>300 °C compared to conventional Al-alloys) with excellent microstructural stability, driven by fine lamellar/rod-like reinforcement (~20 vol%) in an FCC Al matrix and low Ce solubility/diffusivity in Al [22–25]. Their narrow freezing range enables casting [22,23] and fusion-based additive manufacturing [24,25]. Ni-based eutectics exhibit similar strength retention [26–29] but suffer from poor room-temperature ductility due to >50 vol. % intermetallic phases [29], highlighting the importance of eutectic system selection.

This study demonstrates a castable multiphase Nb-based superalloy featuring a (Nb-Mo solid solution)-($Nb_2C$) eutectic microstructure achieving exceptional ultrahigh-temperature strength retention, high thermal stability, coarsening resistance, non-zero room-temperature tensile ductility, low density, and low cost. Oxidation resistance of refractory Nb-based alloys is generally poor but can be addressed via mature silicide-based coating technologies [30].

Alloy design utilized ThermoCalc TCHEA8 database phase diagrams for the Nb-Mo-C system. The Nb-10Mo-9.5C (at. %) alloy (designated NMC-1, denoting the first-generation Nb-Mo-C-based eutectic superalloy) was fabricated by vacuum arc melting high-purity (>99.9%) Nb and Mo with high-purity (>99.5%) graphite, weighed to nominal composition 80.5 at. % Nb, 10 at. % Mo, 9.5 at. % C (87.45 wt. % Nb, 11.22 wt. % Mo, 1.33 wt. % C). Elements were placed in a copper crucible evacuated to 10 mtorr and backfilled with argon twice. After multiple remelts until homogeneous, the alloy was cast into rectangular ingots (82 × 12.5 × 3.8 mm). Density was measured by Archimedes' method (ASTM B96 [31]). Vickers hardness testing employed a 0.5 kgf load. Tensile dogbone specimens were extracted by wire EDM. Room-temperature tensile tests used an MTS 810 servohydraulic system at $1 \times 10^{-3}$ $s^{-1}$ strain rate. Ultrahigh-temperature tensile tests at 1300 and 1500 °C employed a custom Instron 1330/Centorr high-vacuum furnace system ($5 \times 10^{-5}$ torr, ±5 °C) at identical strain rate. Coarsening behavior was assessed via annealing at 1400 °C for 100 hours under high vacuum ($5 \times 10^{-5}$ torr) in Ta foil with Ti getters.

Microstructural characterization utilized SEM (ThermoFisher Quattro S ESEM, Scios DualBeam FIB/SEM, Zeiss Auriga Crossbeam FIB/SEM, ThermoFisher Helios 5 DualBeam Plasma-FIB), EDS, EBSD, XRD (Empyrean, Cu-Kα, 6.66 ° /min), and STEM (JEOL 2100F-AC, 200 kV, Oxford X-MaxN TSR EDS). SEM specimens were mechanically ground (320–1200 grit) then electropolished (10 vol% $H_2SO_4$/methanol, 80 mA, −30 °C) or vibratory polished (0.02 μm colloidal silica, 24 h). STEM specimens were prepared via FIB liftout targeting multiphase regions.

As-cast NMC-1 microstructure (**Fig. 1a–d**) comprises Mo-rich dendrites (350±4 HV) and equiaxed lamellar eutectic colonies (390±2 HV). In the eutectic region, STEM-EDS confirms that the carbide phase is depleted in Mo and enriched in C. The grain size of the eutectic colonies was determined to be 11.3±2.0 µm, the $Nb_2C$ and BCC lamellae widths are 166±31 nm and 284±72 nm, respectively, for a total interlamellar spacing of ~450 nm. Calculated phase diagram (**Fig. 1e**) of NMC-1 indicates a eutectic temperature (melting point) of 2311 ºC and a composition of 9.67 at. % C, with a Nb-Mo solid solution molar phase fraction of ~70% and an $M_2C$ (M = Metal) carbide phase fraction of ~30%. XRD data in **Fig. 1f** verifies the coexistence of 71.2% BCC Nb-Mo solid solution and 28.8% hexagonal $Nb_2C$ carbide.

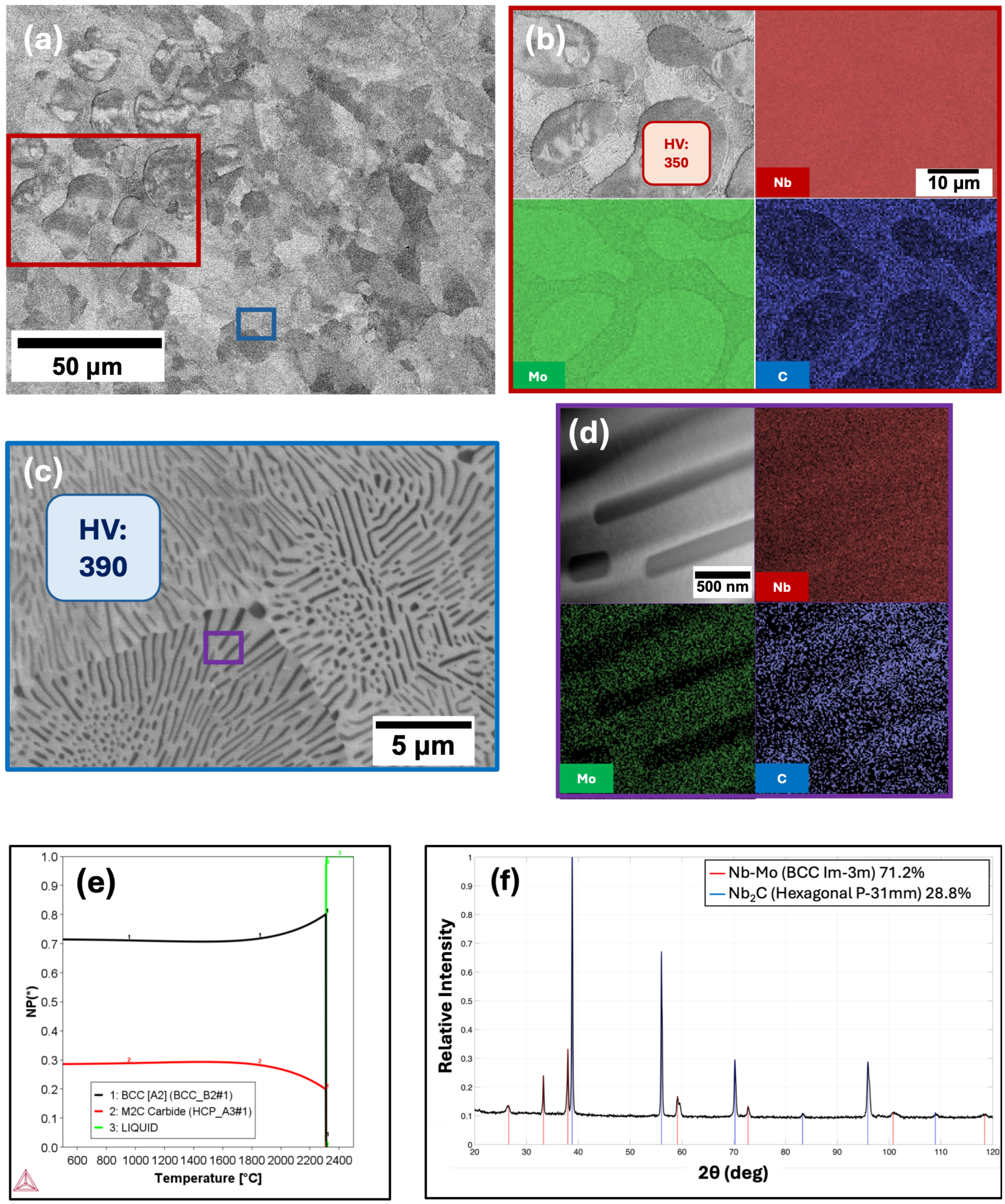

**Fig. 1** (a) SEM micrograph of as-cast NMC-1 showing the eutectic structure. The red inset (b) shows an expanded view of a Mo-rich dendritic region with corresponding EDS mapping. The blue inset (c) shows a region with uniform lamellar eutectic structure and equiaxed grains; Vickers hardness values for both regions are also given. The purple inset (d) gives the STEM-EDS result for the lamellar and interlamellar region. (e) Calculated phase diagram of NMC-1, with the lines indicating the fractions of each phase. (f) XRD data of NMC-1, showing the coexistence of the Nb-Mo solid solution and $Nb_2C$ carbide phase.

The design strategy exploits eutectic phase diagrams between carbon and refractory elements. Using Nb-Mo-C as a proof-of-concept, this approach is extensible to other refractory metal combinations and alternative metalloids (B, N, O) as partial C substitutes. Binary Nb-C and Mo-C phase diagrams (**Fig. S1**) both exhibit BCC+$M_2C$ two-phase fields. The Nb-C eutectic lies closer to the BCC boundary than Mo-C, reducing carbide fraction, which is critical for ductility [29]. However, higher C solubility in BCC Nb raises thermal stability concerns, whereas BCC Mo shows lower C solubility. To combine these favorable attributes, Mo was alloyed into Nb-10C (binary Nb-C eutectic composition). The ternary Nb-Mo-C 10 at. % C isopleth (**Fig. S2**) shows BCC+$M_2C$ stability below melting for Mo contents up to ≥40 at.%. A 10 at. % Mo content was selected to suppress C solubility in the BCC phase (**Fig. S3**) to 1500 °C while limiting Mo-induced embrittlement. The 10 at. % Mo isopleth (**Fig. S3**) indicates eutectic composition slightly below 10 at. % C, yielding final composition Nb-10Mo-9.5C.

NMC-1 exhibits several key attributes. First, the simple composition and processing route enable extremely low feedstock cost (~$85/kg) and density of 8.65 g/cc, yielding favorable specific strength-to-cost. Second, near-zero carbon solubility in the solid solution to 1500 °C (**Figs. 1b, S3b**) ensures nearly all carbon resides in discontinuous $Nb_2C$ lamellae. Because carbon cannot dissolve into the solid solution and diffuse between lamellae, the strengthening phase is protected from coarsening at least up to 1500 °C. This is confirmed by annealing at 1400 °C for 100 hours (**Fig. S4**) showing no $Nb_2C$ coarsening or grain growth. Third, tensile ductility derives from 30 vol% discrete lamellar $Nb_2C$ reinforcement within a continuous, ductile Nb-Mo solid solution [32], with clean grain boundaries free of embrittling phases. Unlike historical Nb-Si superalloys near-

eutectic compositions (limited by low eutectic temperature of 1880 °C, excessive brittle $Nb_5Si_3$ fraction >50%, and coarse $Nb_3Si$ formation) [33,34], NMC-1 overcomes these limitations. While David and Brody [35] previously investigated Nb-$Nb_2C$ eutectic phase stability and growth kinetics without exploring mechanical properties, recent eutectic/hypoeutectic NbMoW-C RCCAs show high compressive strength and thermal stability [36,37] but are expected to exhibit poor tensile ductility due to high Group VI content.

Tensile stress-strain curves (**Fig. 2a**) show NMC-1 yield strengths of 597 MPa, 293 MPa, and 188 MPa, with ultimate tensile strengths (UTS) of 711 MPa, 337 MPa, and 226 MPa at RT, 1300 °C, and 1500 °C, respectively. Room-temperature tensile ductility was ~1% total elongation. The UTS/HV ratio (≈2.4) deviates from the empirical 3× rule-of-thumb, suggesting possible embrittlement mechanisms that caused premature failure. Post-mortem fractography (**Fig. S5**) reveals casting pores and unmelted graphite flakes at fracture surfaces, confirming NMC-1 is intrinsically ductile and free of grain boundary embrittlement, where future processing optimization can further improve ductility.

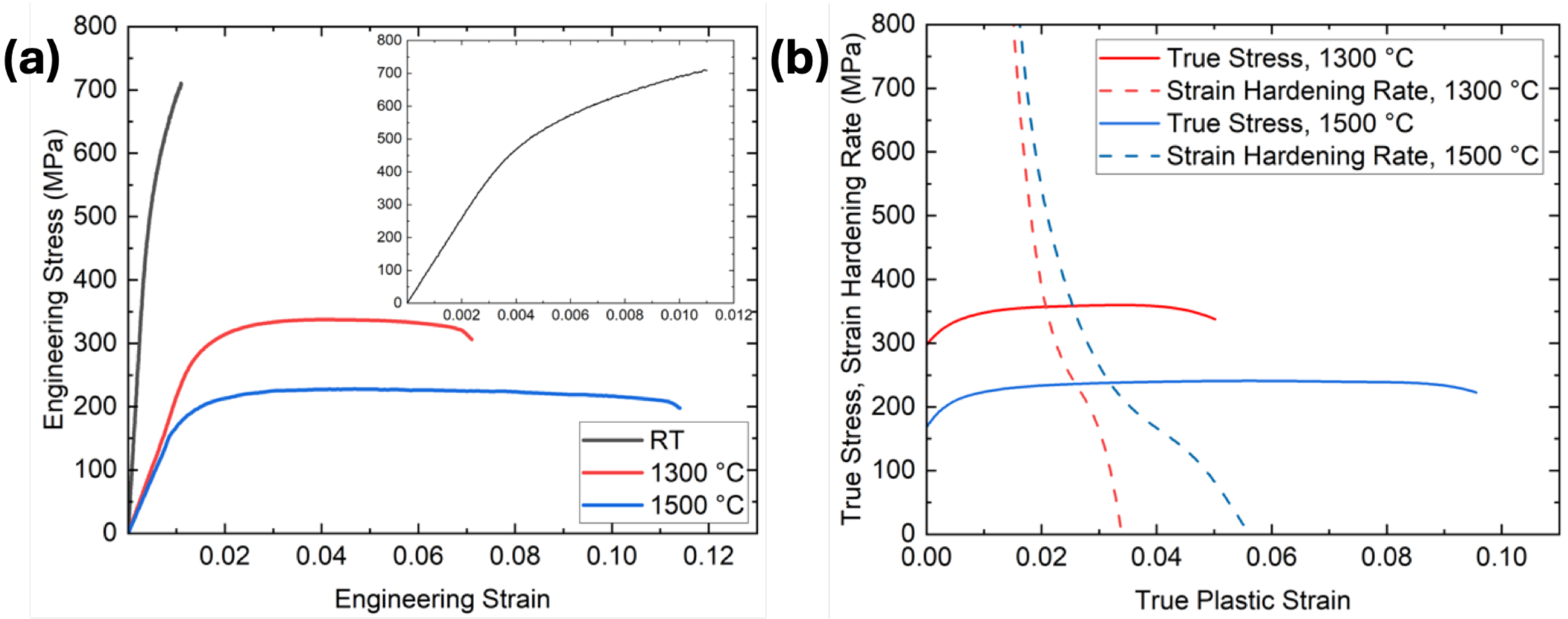

**Fig. 2** (a) Engineering stress-strain curves of NMC-1 at room temperature, 1300 ºC. and 1500 ºC. The inset shows an expanded view of the room-temperature tensile data with a total elongation of ~1%. (b) True stress-true plastic strain curves of NMC-1 along with strain hardening rates $d\sigma/d\epsilon_p$ at 1300 ºC and 1500 ºC.

**Fig. 2b** highlights an even more important attribute of NMC-1's ultrahigh-temperature mechanical performance than its strength values alone: significant strain hardening and extensive uniform elongation (more than 25% of the total elongation) are evident during deformation up to 1500 °C. In conventional refractory alloys and many ductile RCCAs, the yield strength converges with the UTS at these extreme temperatures [38], driven by the onset of dynamic recovery and recrystallization [39], significant grain coarsening [40], and strain-softening mechanisms such as kink band formation [41]. The absence of strain hardening and uniform elongation is detrimental for load-bearing components, where the part can fail abruptly under a sustained load that exceeds the material's yield strength. **Fig. 3a-c** presents the EBSD results for NMC-1 deformed to failure at 1500 °C, showing that the grain structure from the initial microstructure is largely retained after a tensile strain of ~11%. The corresponding kernel average misorientation (KAM) map indicates that a high degree of misorientation is localized at BCC matrix/$Nb_2C$ interfaces and otherwise uniformly distributed throughout the solid solution phase. The high fraction of $Nb_2C$ effectively suppresses the coarsening of the lamellar structure during tensile deformation at 1500 ºC.

Specifically, two classical mechanisms govern eutectic coarsening: (1) continuous coarsening via solid-state diffusion through the matrix phase, or more slowly via diffusion along interphase boundaries, and (2) discontinuous coarsening, nucleated and propagated along colony boundaries [26,28,42–45]. Our material exhibits exceptional resistance to both pathways, particularly discontinuous coarsening, suggesting that colony boundaries maintain low interfacial energy. The coarsening resistance is further enhanced by the exceptionally clean and well-aligned eutectic microstructure, which contains very few defect sites capable of initiating continuous coarsening. This is supported by our EBSD orientation analysis, where the BCC $\{112\}$ orientation is parallel to HCP $\{10\bar{1}1\}$, which is consistent with the low-misfit-strain Burgers orientation relationship widely observed in dual phase Ti alloys [46]. On the other hand, STEM low-angle annular dark field (STEM-LAADF) results in **Fig. 3d-e** indicate ample dislocation activity in the Nb-Mo BCC solid solution and absence of deformation in forms of dislocation activity or shear banding in the $Nb_2C$ phase. The strengthening phase caused significant dislocation pile-up in front of the BCC-$Nb_2C$ interface, as indicated by the yellow arrows.

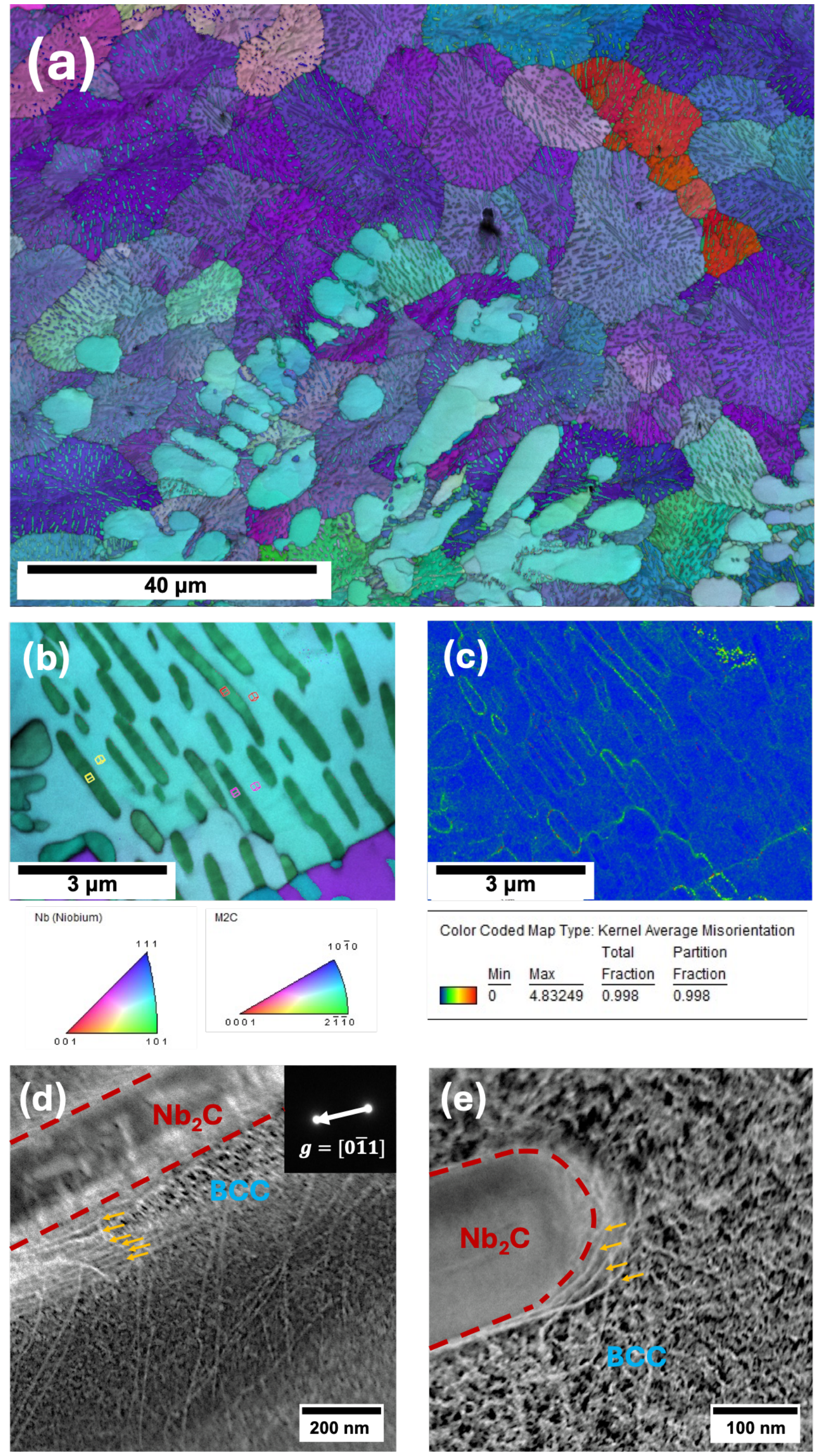

**Fig. 3** (a) EBSD-inverse pole figure (IPF) color map for the overall microstructure of NMC-1 deformed at 1500 ºC to failure. (b) Expanded view in the lamellar region, where the crystallographic orientation of {112} BCC matrix and {10-10} HCP Nb2C are demarcated. (c) KAM map for the deformed lamellar region. (d) STEM-LAADF micrograph showing dislocation activity after deformimg at 1500 ºC to failure in the BCC matrix imaged under [011] zone axis along $g = [0\bar{1}1]$ two-beam condition. (e) Expanded view of a $Nb_2C$ lamella terminated within the BCC matrix with dislocations wrapping along the interface. Dislocation pile-up again the matrix-lamella interface is indicated in yellow arrows.

The mechanical properties of NMC-1 are compared with commercial Nb alloys [3,5,6] and recently developed Nb-rich ductile RCCAs [41,47] in **Fig. 4a-b**. It is evident that the RT yield strength and UTS of NMC-1 is comparable to, or often lower than, those of Nb-rich RCCAs and AM Nb alloys. These alloys are strengthened, respectively, by concentrated solid-solution strengthening from high-misfit elements (Hf and V) alloyed with Nb, and by the high defect density introduced during additive manufacturing (AM). It is noted that the reported UTS of NMC-1 at RT may be undervalued, as embrittlement caused by casting defects likely prevents the alloy from exhibiting its full strain-hardening behavior. Up to 1200 °C, wrought Nb521 and AM Nb alloys show excellent strength retention, which can be attributed to strengthening from a small fraction of carbide and oxide dispersoids in the matrix, either intentionally added or formed during the AM process [6]. However, at ultrahigh temperatures beyond 1200 °C, conventional Nb alloys and ductile RCCAs exhibit significant strength loss, indicating that the strengthening contributions from AM, dispersion strengthening by a small volume fraction of carbides and oxides, and concentrated solid-solution strengthening all diminish substantially. In contrast, NMC-1 retains half of its RT strength at 1300 °C and one third of its RT strength at 1500 °C. These data place NMC-1 favorably in terms of ultrahigh temperature mechanical properties compared to existing Nb alloys [4]. The prototype alloy exhibits UTS approximately 200 % greater than C-103 and 30% greater than Nb521 at 1300 °C, increasing to 300 % and 100 % greater than C-103 and Nb521, respectively, at 1500 °C. These gains are attributed to the high volume fraction of the strengthening phase and the substantial Mo content, which collectively raise the elastic modulus and suppress diffusivity-controlled deformation mechanisms [48].

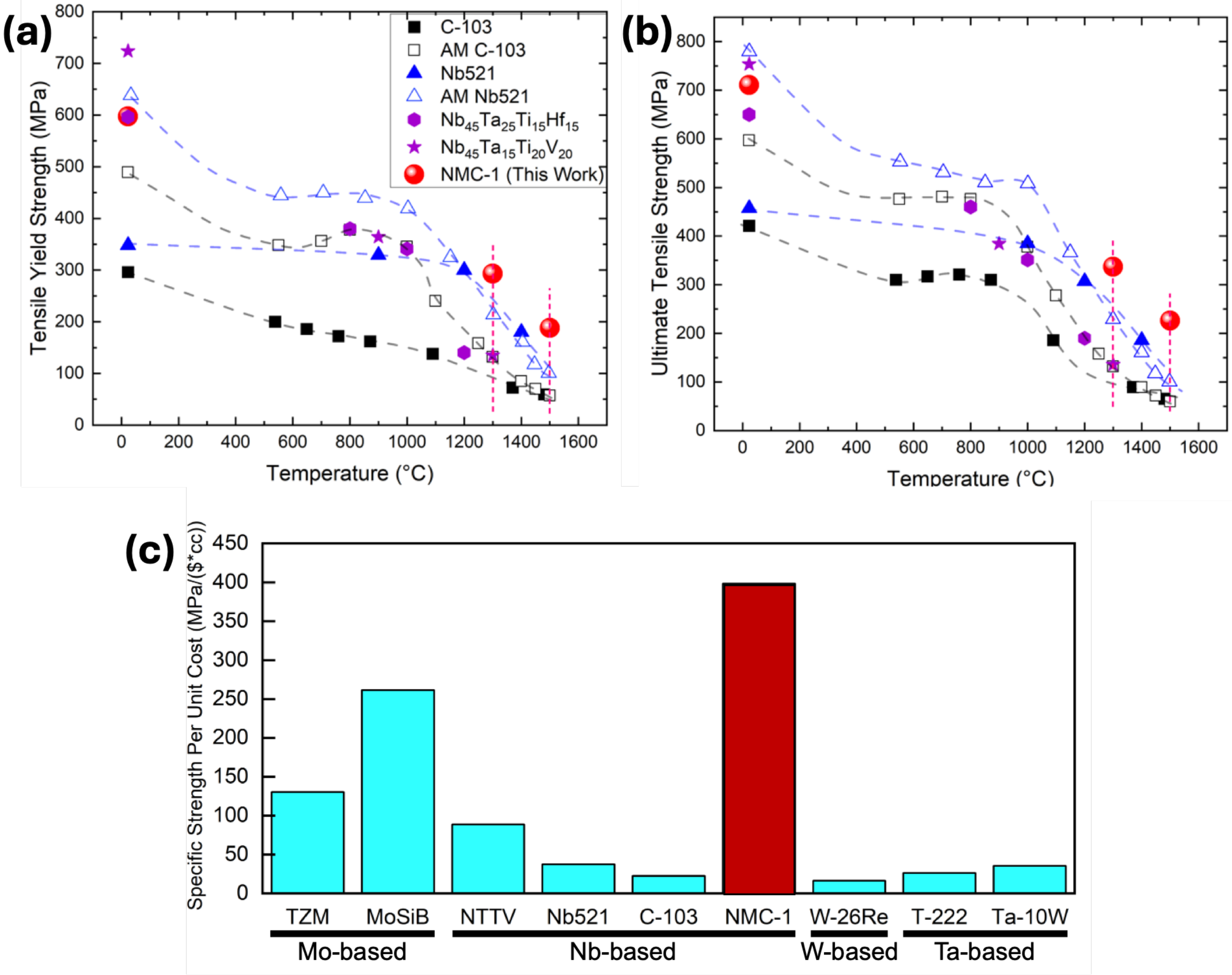

**Fig. 4** (a) 0.2% offset tensile yield strengths and (b) ultimate tensile strengths of wrought and additively manufactured (AM) C-103, wrought and AM Nb521, $Nb_{45}Ta_{25}Ti_{15}Hf_{15}$, $Nb_{45}Ta_{15}Ti_{20}V_{20}$, and NMC-1 developed in this work. (c) gives the specific strength per unit cost at 1300 ºC for a variety of Mo-, Nb-, W-, and Ta-based alloys.

From an application perspective spanning aerospace propulsion and re-entry vehicles, NMC-1 is compared against a wider range of candidate refractory systems: ductile commercial Ta-, Mo- and W-based refractory alloys [47,49–53]. Using Ashby's material selection method [54], a material index of yield strength normalized by density and unit cost (per mass) is employed at an operating temperature of 1300 °C. Wrought commercial alloys were selected where available for their lower cost; powder-based alloys were used only when necessary (primarily dispersion-strengthened systems). The comparison in **Fig. 4c** evaluates TZM (Mo-0.5Ti-0.08Zr, in wt.%), MoSiB (Mo-

12Si-8.5B, in at. %), non-equiatomic NTTV ($Nb_{45}Ta_{25}Ti_{20}V_{20}$, in at. %), Nb521, C-103, W-26Re (in wt. %), T-222 (Ta-10W-2.5Hf-0.01C, in wt. %) and Ta-10W (in wt. %) as well as NMC-1. Note that the alloys' compositions are expressed using a mixture of weight and atomic percentages in order to preserve their established naming conventions. NMC-1 represents a major advancement in ultrahigh-temperature structural materials: it combines the low density of Nb alloys with significantly improved high-temperature strength while maintaining extremely low cost. Mo alloys can outperform commercial Nb alloys due to higher strength and lower cost (albeit slightly higher density), but NMC-1 demonstrates clear superiority. W-Re and Ta alloys are uncompetitive due to high weight and cost, justified only above ~2000 °C where Nb- and Mo-based alloys begin to fail. Here, we chart a course for future studies to further improve NMC-1 in several aspects: optimizing solidification processing to alleviate casting defects and unlock greater tensile ductility; further alloying with W, Hf, and V to leverage additional solid-solution strengthening; exploring higher cooling rates, including laser-based additive manufacturing, to achieve a finer nanolamellar structure to improve strength and toughness; and investigating interfacial behavior during ultrahigh-temperature deformation, with interfacial engineering (e.g., using elements such as boron) to improve interface stability.

In summary, a novel castable eutectic Nb-based superalloy, Nb-10Mo-9.5C (NMC-1), was developed, forming a lamellar microstructure of ~70% Nb-Mo BCC solid solution and ~30% $Nb_2C$ carbide. NMC-1 retains half its RT strength at 1300 °C and one-third at 1500 °C, outperforming C-103, Nb521, and WC-3009 by 50-200% over this range, while sustaining strain hardening and uniform elongation up to 1500 °C. Coarsening of the eutectic structure was confirmed to be minimal after annealing at 1400 °C for 100 hours. After ultrahigh-temperature deformation, STEM reveals dislocation activities in the BCC matrix coupled with dislocation pinning at $Nb_2C$ boundaries. RT ductility (~1%) is currently limited by casting defects rather than intrinsic embrittlement, leaving clear room for improvement. With a density of 8.65 g/cc and cost of ~$85/kg, NMC-1 outperforms commercial Nb, Mo, Ta, and W alloys on a specific strength-per-cost basis at 1300 °C. These results establish NMC-1, and the underlying eutectic carbide-strengthening strategy, as a promising and extensible paradigm for future ultrahigh-temperature structural alloy design.

**Acknowledgements**

Research at UC Davis was supported by discretionary funding provided by the Department of Materials Science and Engineering. A part of this study was carried out at the UC Davis Center for Nano- and Micro-Manufacturing (CNM2) and the Advanced Material Characterization and Testing (AMCaT) Facility. Funding for the Thermo Fisher Quattro S was provided by the National Science Foundation Grant No. MRI-1725618.

Research at UT Knoxville was supported primarily by discretionary funds available to EAL. EAL also acknowledges partial support through the UTK's National Science Foundation Material Research Science and Engineering Center (NSF-MRSEC) Center for Advanced Materials and Manufacturing (DMR-2309083). SEM characterization at UTK was performed at the UT Institute for Advanced Materials and Manufacturing (IAMM) Electron Microscopy Center, while X-ray diffraction analysis was performed at IAMM's Diffraction Facility.

## References

[1] J.H. Perepezko, The Hotter the Engine, the Better, Science 326 (2009) 1068–1069. https://doi.org/10.1126/science.1179327.

[2] T.M. Pollock, S. Tin, Nickel-based superalloys for advanced turbine engines: chemistry, microstructure and properties, J. Propuls. Power 22 (2006) 361–374.

[3] Ltd. Songhan Plastic Technology Co., ATI Wah Chang Nb/Nb Alloy C-103 Datasheet, (2026). https://www.lookpolymers.com/polymer_ATI-Wah-Chang-NbNb-Alloy-C-103.php.

[4] C.C. Wojcik, W. Chang, Thermomechanical processing and properties of niobium alloys., in: 2001: pp. 163–173.

[5]: 朱宝辉, : 吴向东, : 万敏, : 赵刚, : 曹艳飞, : 罗文, : 李树荣, : 何季麟, 航天用高温铌合金研究进展, 中国有色金属学报 33 (2023) 1–26. https://doi.org/10.11817/j.ysxb.1004.0609.2022-42910.

[6] E. Brizes, J. Milner, A Comparison of Niobium Alloys C103 and Nb521, in: 2025.

[7] O.N. Senkov, D.B. Miracle, K.J. Chaput, J.-P. Couzinie, Development and exploration of refractory high entropy alloys—A review, J. Mater. Res. 33 (2018) 3092–3128. https://doi.org/10.1557/jmr.2018.153.

[8] D.B. Miracle, O.N. Senkov, C. Frey, S. Rao, T.M. Pollock, Strength vs temperature for refractory complex concentrated alloys (RCCAs): A critical comparison with refractory BCC elements and dilute alloys, Acta Mater. 266 (2024) 119692. https://doi.org/10.1016/j.actamat.2024.119692.

[9] O.N. Senkov, G.B. Wilks, D.B. Miracle, C.P. Chuang, P.K. Liaw, Refractory high-entropy alloys, Intermetallics 18 (2010) 1758–1765. https://doi.org/10.1016/j.intermet.2010.05.014.

[10] O.N. Senkov, G.B. Wilks, J.M. Scott, D.B. Miracle, Mechanical properties of Nb25Mo25Ta25W25 and V20Nb20Mo20Ta20W20 refractory high entropy alloys, Intermetallics 19 (2011) 698–706. https://doi.org/10.1016/j.intermet.2011.01.004.

[11] P. Kumar, X. Gou, D.H. Cook, M.I. Payne, N.J. Morrison, W. Wang, M. Zhang, M. Asta, A.M. Minor, R. Cao, Y. Li, R.O. Ritchie, Degradation of the mechanical properties of NbMoTaW refractory high-entropy alloy in tension, Acta Mater. 279 (2024) 120297. https://doi.org/10.1016/j.actamat.2024.120297.

[12] O.N. Senkov, J.M. Scott, S.V. Senkova, D.B. Miracle, C.F. Woodward, Microstructure and room temperature properties of a high-entropy TaNbHfZrTi alloy, J. Alloys Compd. 509 (2011) 6043–6048. https://doi.org/10.1016/j.jallcom.2011.02.171.

[13] O.N. Senkov, D. Isheim, D.N. Seidman, A.L. Pilchak, Development of a Refractory High Entropy Superalloy, Entropy 18 (2016) 102. https://doi.org/10.3390/e18030102.

[14] D.B. Miracle, M.-H. Tsai, O.N. Senkov, V. Soni, R. Banerjee, Refractory high entropy superalloys (RSAs), Scr. Mater. 187 (2020) 445–452. https://doi.org/10.1016/j.scriptamat.2020.06.048.

[15] E.A. Lass, On the Thermodynamics and Phase Transformation Pathways in BCC-B2 Refractory Compositionally Complex Superalloys, Metall. Mater. Trans. A 53 (2022) 4481–4498. https://doi.org/10.1007/s11661-022-06844-6.

[16] M.K. Moczadlo, E.A. Lass, Microstructure and Phase Equilibria in BCC-B2 Nb-Ti-Ru Refractory Superalloys, Materials 17 (2024) 5429. https://doi.org/10.3390/ma17225429.

[17] P. Kumar, S.J. Kim, Q. Yu, J. Ell, M. Zhang, Y. Yang, J.Y. Kim, H.-K. Park, A.M. Minor, E.S. Park, R.O. Ritchie, Compressive vs. tensile yield and fracture toughness behavior of a body-centered cubic refractory high-entropy superalloy

Al0.5Nb1.25Ta1.25TiZr at temperatures from ambient to 1200°C, Acta Mater. 245 (2023) 118620. https://doi.org/10.1016/j.actamat.2022.118620.
[18] S.A. Kube, C. Frey, C. McMullin, B. Neuman, K.M. Mullin, T.M. Pollock, Navigating the BCC-B2 refractory alloy space: Stability and thermal processing with Ru-B2 precipitates, Acta Mater. 265 (2024) 119628. https://doi.org/10.1016/j.actamat.2023.119628.
[19] C. Frey, H. You, S. Kube, G.H. Balbus, K. Mullin, S. Oppenheimer, C.S. Holgate, T.M. Pollock, High Temperature B2 Precipitation in Ru-Containing Refractory Multi-principal Element Alloys, Metall. Mater. Trans. A 55 (2024) 1739–1764. https://doi.org/10.1007/s11661-024-07368-x.
[20] C. Frey, B. Neuman, K. Mullin, A. Botros, J. Lamb, C.S. Holgate, S.A. Kube, T.M. Pollock, On the stability of coherent HfRu- and ZrRu-B2 precipitates in Nb-based alloys, Mater. Des. 247 (2024) 113385. https://doi.org/10.1016/j.matdes.2024.113385.
[21] B.J. Crossman, J. Wang, L. Perrière, S.A. Chen, J.-P. Couzinié, M. Ghazisaeidi, M.J. Mills, Multi-modal characterization of the B2 phase in the Ta-Re binary system, Acta Mater. 293 (2025) 121097. https://doi.org/10.1016/j.actamat.2025.121097.
[22] Z.C. Sims, D. Weiss, S.K. McCall, M.A. McGuire, R.T. Ott, T. Geer, O. Rios, P.A.E. Turchi, Cerium-Based, Intermetallic-Strengthened Aluminum Casting Alloy: High-Volume Co-product Development, JOM 68 (2016) 1940–1947. https://doi.org/10.1007/s11837-016-1943-9.
[23] D. Weiss, Improved High-Temperature Aluminum Alloys Containing Cerium, J. Mater. Eng. Perform. 28 (2019) 1903–1908. https://doi.org/10.1007/s11665-019-3884-2.
[24] A. Plotkowski, O. Rios, N. Sridharan, Z. Sims, K. Unocic, R.T. Ott, R.R. Dehoff, S.S. Babu, Evaluation of an Al-Ce alloy for laser additive manufacturing, Acta Mater. 126 (2017) 507–519. https://doi.org/10.1016/j.actamat.2016.12.065.
[25] K. Sisco, A. Plotkowski, Y. Yang, D. Leonard, B. Stump, P. Nandwana, R.R. Dehoff, S.S. Babu, Microstructure and properties of additively manufactured Al–Ce–Mg alloys, Sci. Rep. 11 (2021) 6953. https://doi.org/10.1038/s41598-021-86370-4.
[26] S.B. Haider, E. Heon, M. Neveau, P. Chen, A. Houston, O. Rios, E.A. Lass, Castable eutectic Ni–Ce high temperature alloys strengthened by γ/γ′ microstructure, J. Mater. Res. Technol. 28 (2024) 3943–3950. https://doi.org/10.1016/j.jmrt.2023.12.263.
[27] C.S. Tiwary, A. Kashiwar, S. Bhowmick, K.C. Hari Kumar, K. Chattopadhyay, D. Banerjee, Engineering an ultrafine intermetallic eutectic ternary alloy for high strength and high temperature applications, Scr. Mater. 157 (2018) 67–71. https://doi.org/10.1016/j.scriptamat.2018.07.036.
[28] S. Bushra Haider, I.K. Robin, E.A. Lass, Thermal stability and coarsening of eutectic and near-eutectic Ni–Ce alloys, Intermetallics 174 (2024) 108458. https://doi.org/10.1016/j.intermet.2024.108458.
[29] S.B. Haider, I.K. Robin, E.A. Lass, Phase Equilibria and Microstructure Evolution in the Ni-rich Region of Binary Ni-Ce and Ternary Ni-Ce-X (X = Al, Nb, Cr, Ti) Systems at 900 °C, J. Phase Equilibria Diffus. 46 (2025) 333–349. https://doi.org/10.1007/s11669-025-01196-1.
[30] B.A. Pinto, A.S.C.M. d’Oliveira, Nb silicide coatings processed by double pack cementation: Formation mechanisms and stability, Surf. Coat. Technol. 409 (2021) 126913. https://doi.org/10.1016/j.surfcoat.2021.126913.
[31] ASTM International, Standard Test Methods for Density of Compacted or Sintered Powder Metallurgy (PM) Products Using Archimedes’ Principle, (2023).

https://www.astm.org/Standards/B962.htm.
[32] R.T. Begley, J.H. Bechtold, Effect of alloying on the mechanical properties of niobium, J. Common Met. 3 (1961) 1–12. https://doi.org/10.1016/0022-5088(61)90037-6.
[33] M.E. Schlesinger, H. Okamoto, A.B. Gokhale, R. Abbaschian, The Nb-Si (Niobium-Silicon) system, J. Phase Equilibria 14 (1993) 502–509. https://doi.org/10.1007/BF02671971.
[34] B.P. Bewlay, M.R. Jackson, J.-C. Zhao, P.R. Subramanian, M.G. Mendiratta, J.J. Lewandowski, Ultrahigh-Temperature Nb-Silicide-Based Composites, MRS Bull. 28 (2003) 646–653. https://doi.org/10.1557/mrs2003.192.
[35] S.A. David, H.D. Brody, Growth of niobium-niobium carbide (Nb2C) eutectic and hypereutectic composites by zone melting, Metall. Trans. 5 (1974) 2309–2316. https://doi.org/10.1007/BF02644011.
[36] Y. Zhang, Q. Wei, P. Xie, X. Xu, An ultrastrong niobium alloy enabled by refractory carbide and eutectic structure, Mater. Res. Lett. 11 (2023) 169–178. https://doi.org/10.1080/21663831.2022.2133977.
[37] Q. Shen, X. Wu, Q. Wei, J. Zhang, J. Xu, X. Wang, G. Luo, A strong-ductile niobium alloy enhanced by eutectic Nb2C, Mater. Sci. Eng. A 882 (2023) 145448. https://doi.org/10.1016/j.msea.2023.145448.
[38] J. Wadsworth, T.G. Nieh, J.J. Stephens, Recent advances in aerospace refractory metal alloys, Int. Mater. Rev. 33 (1988) 131–150. https://doi.org/10.1179/imr.1988.33.1.131.
[39] L.H. Mills, M.G. Emigh, C.H. Frey, N.R. Philips, S.P. Murray, J. Shin, D.S. Gianola, T.M. Pollock, Temperature-dependent tensile behavior of the HfNbTaTiZr multi-principal element alloy, Acta Mater. 245 (2023) 118618. https://doi.org/10.1016/j.actamat.2022.118618.
[40] S.I.A. Jalali, K.J. Hemker, Measuring the Mechanical Response of Materials at Extreme Temperatures with Localized Heating, Annu. Rev. Mater. Res. 56 (2026) 283–309. https://doi.org/https://doi.org/10.1146/annurev-matsci-072924-115933.
[41] D.H. Cook, P. Kumar, M.I. Payne, C.H. Belcher, P. Borges, W. Wang, F. Walsh, Z. Li, A. Devaraj, M. Zhang, M. Asta, A.M. Minor, E.J. Lavernia, D. Apelian, R.O. Ritchie, Kink bands promote exceptional fracture resistance in a NbTaTiHf refractory medium-entropy alloy, Science 384 (2024) 178–184. https://doi.org/10.1126/science.adn2428.
[42] A.J. Ardell, Microstructural stability at elevated temperatures, J. Eur. Ceram. Soc. 19 (1999) 2217–2231. https://doi.org/10.1016/S0955-2219(99)00094-1.
[43] K. Jung, H. Conrad, Microstructure coarsening during static annealing of 60Sn40Pb solder joints: II eutectic coarsening kinetics, J. Electron. Mater. 30 (2001) 1303–1307. https://doi.org/10.1007/s11664-001-0115-y.
[44] L. Fu, J. He, S. Lu, Y. Sun, D. Zhu, Y. Mao, Coarsening kinetics of lamellar and equiaxed microstructures of eutectic Au–20Sn during the annealing, J. Mater. Res. Technol. 17 (2022) 2134–2144. https://doi.org/10.1016/j.jmrt.2022.01.162.
[45] X. Li, F. Bottler, R. Spatschek, A. Schmitt, M. Heilmaier, F. Stein, Coarsening kinetics of lamellar microstructures: Experiments and simulations on a fully-lamellar Fe-Al in situ composite, Acta Mater. 127 (2017) 230–243. https://doi.org/10.1016/j.actamat.2017.01.041.
[46] D. He, J.C. Zhu, S. Zaefferer, D. Raabe, Y. Liu, Z.L. Lai, X.W. Yang, Influences of deformation strain, strain rate and cooling rate on the Burgers orientation relationship and variants morphology during β→α phase transformation in a near α titanium alloy, Mater. Sci. Eng. A 549 (2012) 20–29. https://doi.org/10.1016/j.msea.2012.03.110.

[47] A.M. Nahin, J. Pustelnik, J. Dong, T. Zakia, M. Zhang, A low-cost, strong, and ductile single-phase Nb-based refractory complex concentrated alloy, Scr. Mater. 284 (2026) 117488. https://doi.org/10.1016/j.scriptamat.2026.117488.
[48] S. Prasad, A. Paul, Diffusion Parameters in the Nb-Mo System: Revisited, Metall. Mater. Trans. A 40 (2009) 1512–1514. https://doi.org/10.1007/s11661-009-9861-x.
[49] Chinatungsten, TZM Alloy (Titanium Zirconium Molybdenum)-Properties, (n.d.). http://titanium-zirconium-molybdenum.com/TZM-Alloy-Properties.html.
[50] J. Wang, W. Wang, B. Li, R. Li, C. Wang, J. Zhang, T. Wang, G. Zhang, Understanding improved high temperature strength of Mo-Si-B alloy at 1100–1300 °C based on ZrB2 addition, Mater. Charact. 225 (2025) 115126. https://doi.org/10.1016/j.matchar.2025.115126.
[51] E. Lassner, W.-D. Schubert, Tungsten: properties, chemistry, technology of the element, alloys, and chemical compounds, Kluwer Academic/Plenum Publishers New York, 1999.
[52] B. Chen, F. Zheng, J. Li, M. Xue, A. Cui, X. Luo, S. Li, X. Ding, J. Sun, Development of Ductile Refractory Alloys With Excellent High-Temperature Strength for Extreme Environments, Adv. Mater. 38 (2026) e15947. https://doi.org/10.1002/adma.202515947.
[53] J. Conway, Mechanical and physical properties of refractory metals and alloys, 1984.
[54] M.F. Ashby, K. Johnson, Materials and design: the art and science of material selection in product design, Butterworth-Heinemann, 2013.

# Supplementary Materials

**for**

**A Novel Eutectic Nb-Based Superalloy with Exceptional Ultrahigh-Temperature Mechanical Properties**

Ayeman M. Nahin[1, †], Alfredo Navarrete[2, †], Xiaokun Yang[1], Eric A. Lass[2, *], Mingwei Zhang[1, *]

[1]Department of Materials Science and Engineering, University of California, One Shields Ave., Davis, CA 95616, USA

[2]Department of Materials Science and Engineering, University of Tennessee, Knoxville, TN 37996, USA

[†] These authors contribute equally to this paper.

*Corresponding Authors:

Eric A. Lass, Associate Professor

Department of Materials Science and Engineering

University of Tennessee, Knoxville

Email: elass@utk.edu

Mingwei Zhang, Assistant Professor

Department of Materials Science and Engineering

University of California, Davis

Email: mwwzhang@ucdavis.edu

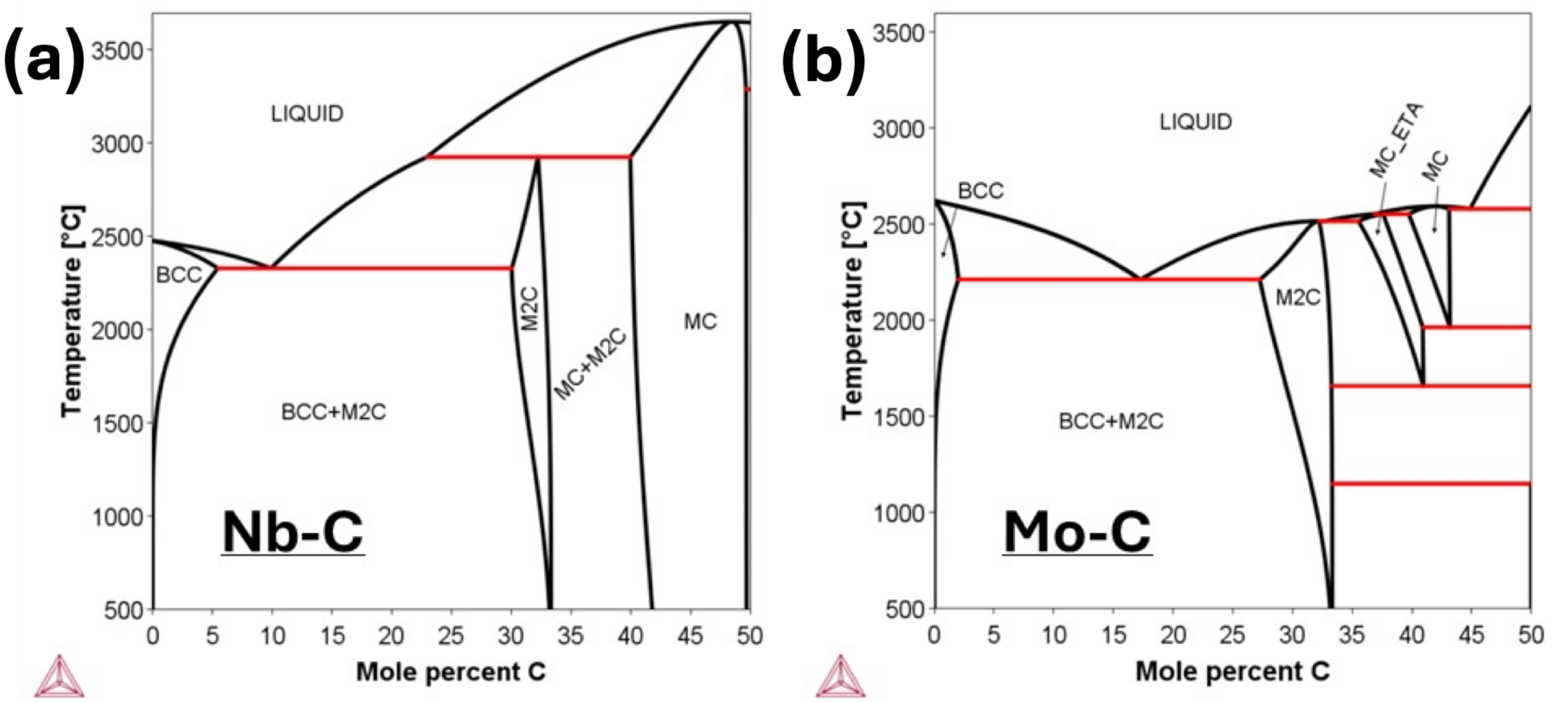

**Fig.S1** Binary (a) Nb-C and (b) Mo-C phase diagrams.

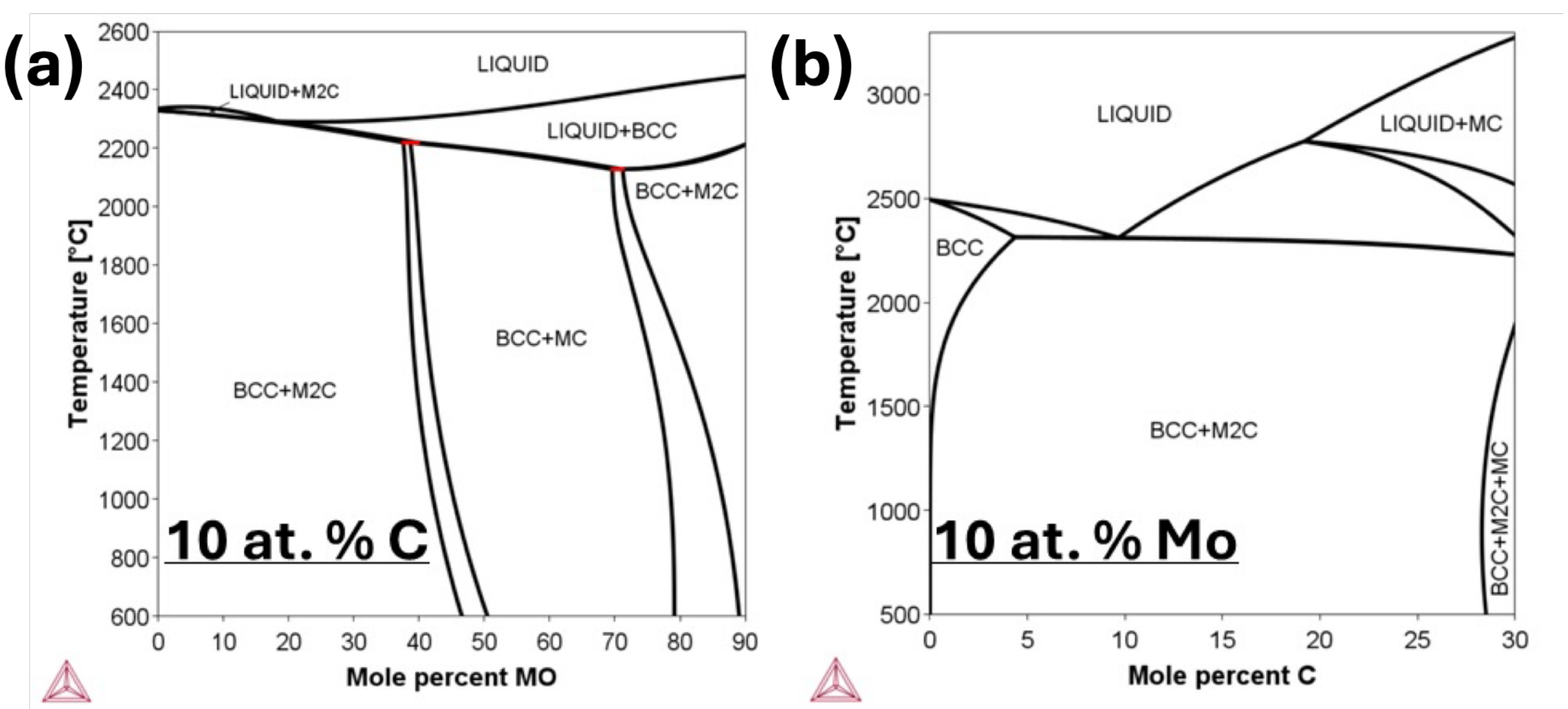

**Fig. S2** Isopleth sections of the ternary Nb-Mo-C system for (a) 10 at. % C and (b) 10 at. % Mo.

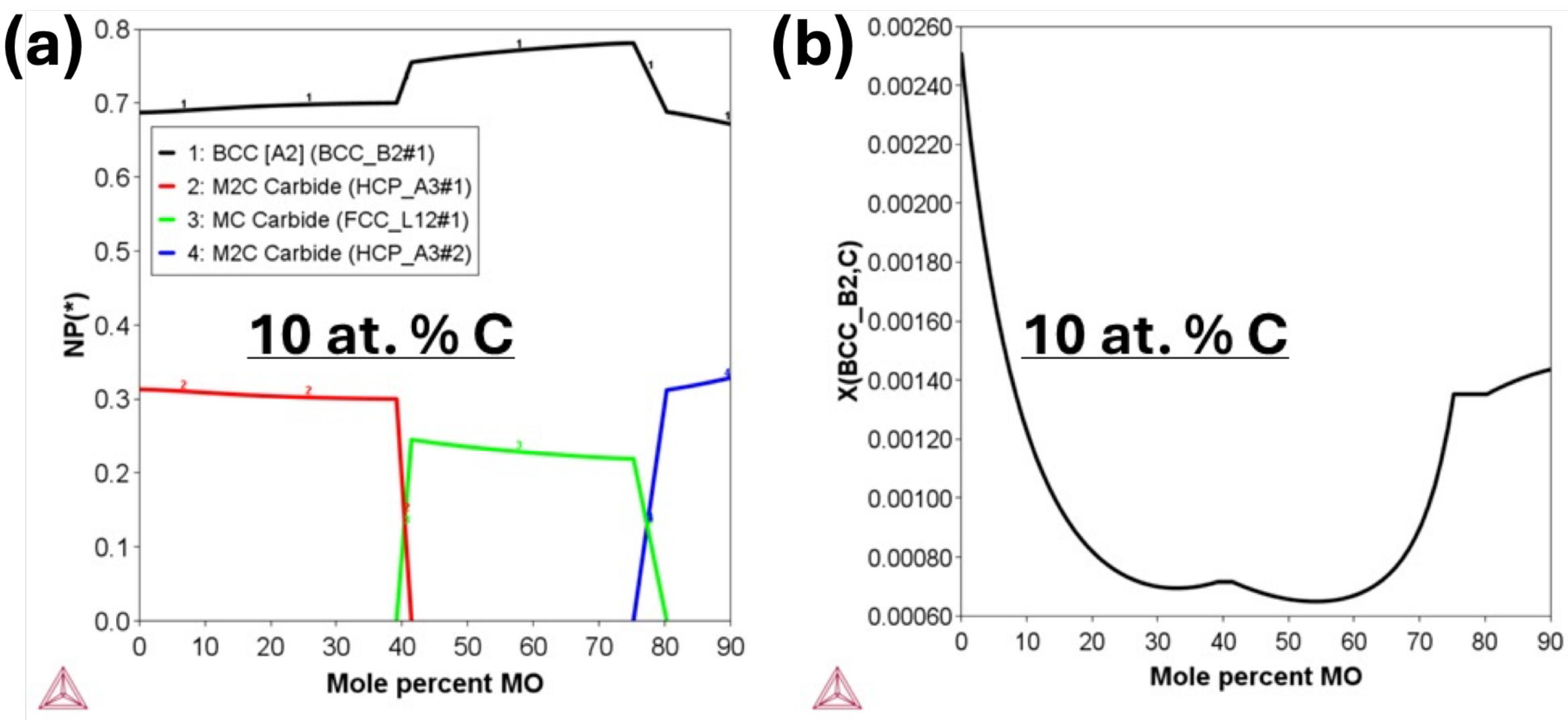

**Fig. S3** (a) Phase fractions of BCC and carbide phases of Nb-10 at. % C as a function of Mo content at 1500 ºC. (b) C solubility in the solid solution as a function of Mo content at 1500 ºC.

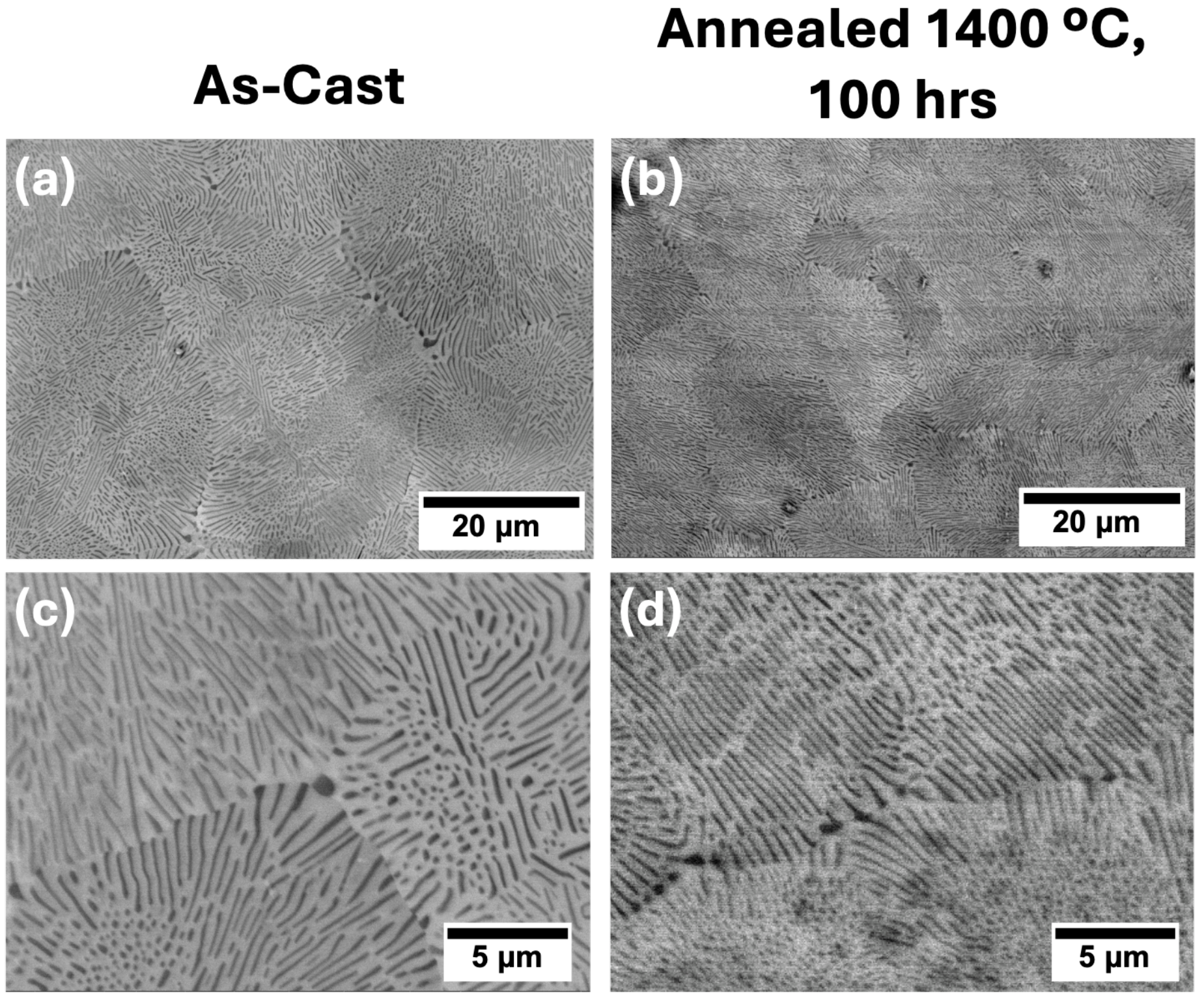

**Fig. S4** SEM micrographs of NMC-1 (a) under as-cast condition; (b) after annealing for 1400 ºC under vacuum for 1000 hours, showing no apparent grain growth. Expanded views of the microstructure is shown for the (c) as-cast; and (d) material annealed at 1400 ºC for 1000 hours, showing no coarsening of $Nb_2C$ lamellae.

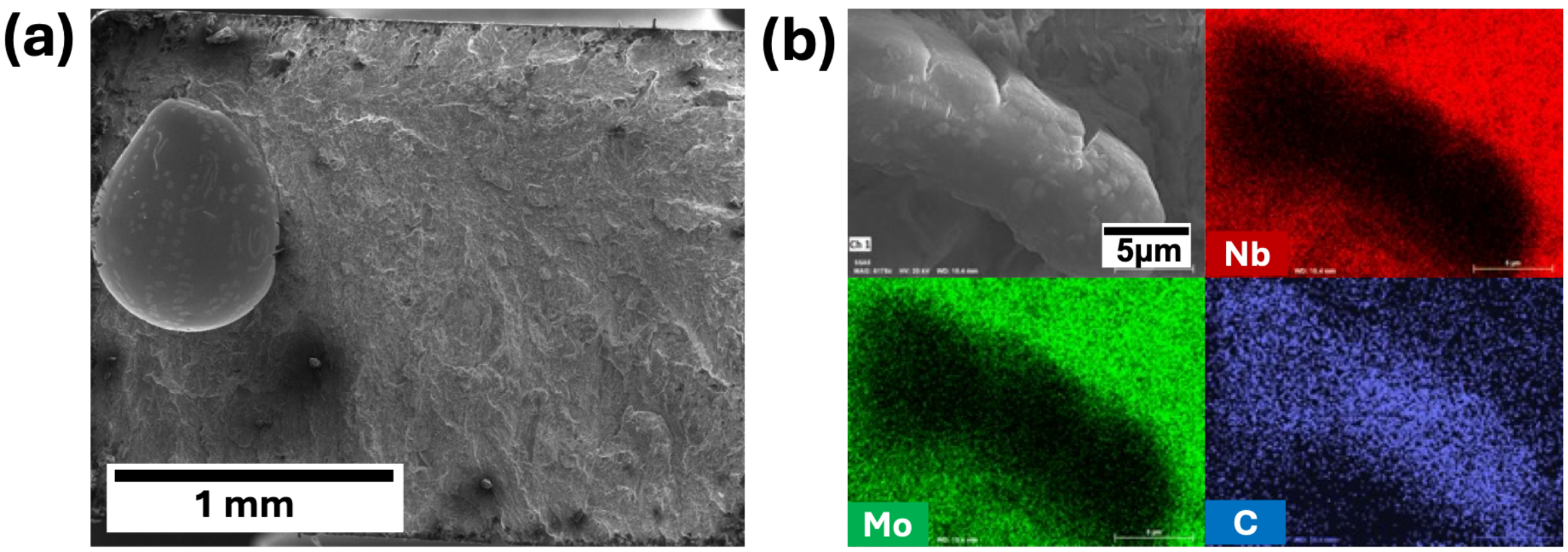

**Fig. S5** (a) SEM micrograph of the fracture surface of NMC-1 after RT tensile testing. A casting pore with a diameter of ~0.5 mm is evident. (b) SEM micrograph and corresponding EDS mapping of an unmelted graphite